\documentclass{jnmp}

\usepackage{amsmath}

\newtheorem{theorem}{Theorem}
\newtheorem{lemma}{Lemma}
\newtheorem*{corollary}{Corollary}

\theoremstyle{definition}
\newtheorem*{definition}{Definition}
\newtheorem{remark}{Remark}

\def\toro{{\mathbb T}}
\def\R{{\mathbb R}}
\def\Z{{\mathbb Z}}
\def\^#1{{\widehat #1}}
\def\wt#1{{\widetilde #1}}

\begin{document}

\renewcommand{\evenhead}{G~Gaeta}
\renewcommand{\oddhead}{The Poincar\'e--Nekhoroshev Map}

\thispagestyle{empty}

\FirstPageHead{10}{1}{2003}{\pageref{gaeta-firstpage}--\pageref{gaeta-lastpage}}{Article}

\copyrightnote{2003}{G~Gaeta}

\Name{The Poincar\'e--Nekhoroshev Map}
\label{gaeta-firstpage}

\Author{Giuseppe GAETA}

\Address{Dipartimento di Matematica, Universit\`a di Milano,
I-20133 Milano, Italy\\
 Dipartimento di Fisica, Universit\`a di Roma,
I--00185 Roma, Italy \\
E-mail: g.gaeta@tiscali.it, gaeta@roma1.infn.it}

\Date{Received March 04, 2002;
Accepted June 28, 2002}

\begin{abstract}
\noindent
We study a generalization of the familiar Poincar\'e
map, first implicitely introduced by N~N~Nekhoroshev
in his study of persistence of invariant tori in hamiltonian systems,
and discuss some of its properties and applications. In particular,
we apply it to study persistence and bifurcation of invariant tori.
\end{abstract}

\section*{Introduction}

The {\it Poincar\'e map } is a classical tool in the study of a
dynamical system around a known periodic solution (see e.g.\
\cite{Arn2,Dem,Gle,GuH,Rue,Ver}).

Here we want to study a dynamical system around a multi-periodic
solution (i.e.\ an invariant torus $\toro^k$, $k > 1$):
in this case the Poincar\'e map has several drawbacks, and it would
be more convenient to somehow quotient out the degrees of freedom
corresponding to motion along the invariant torus (and
transversal to the dynamical flow). However, such a quotient is
in general terms ill-defined out of the torus itself.

In a paper \cite{Nek} devoted to persistence of invariant tori in partially integrable
hamiltonian systems with $n$ degrees of freedom and $k$ integrals in involution
($1 < k < n$), N~N~Nekhoroshev
devised a way to overcome this obstacle,
and generalized the classical  Poincar\'e--Lyapounov theorem.
The main nondegeneracy condition for this theorem was expressed
in terms of monodromy operators. Unfortunately, his discussion
was very short and somehow not easy reading, and -- as far as I know --
he never published a proof of this result.

Here we note that the whole matter is better understood
in terms of a generalization of the Poincar\'e map, which
in my opinion is implicitely introduced in \cite{Nek} and which
I~will call the Poin\-ca\-r\'e--Ne\-kho\-ro\-shev map.
This map is of interest {\it per se}, i.e.\
not just for the Poincar\'e--Lyapounov--Nekhoroshev theorem.

The aim of the present note is to discuss in detail the
Poin\-ca\-r\'e--Ne\-kho\-ro\-shev map (which can be defined also for
non-hamiltonian systems), its geometry and its spectrum. In particular
I will discuss how this bypasses the obstruction to considering
a symmetry quotient (note that when such obstruction is not present,
we can pass to the quotient system and apply standard Poincar\'e theory there),
i.e.\ how this can be applied without assuming regularity of global
invariant manifolds near the invariant torus, and its relation with monodromy.

The {\bf plan of the paper} is as follows.
In Section 1 we fix some basic notation and recall background results
concerning closed trajectories on tori. In Section 2 we
recall the definition of the standard Poincar\'e map,
and provide a geometrical interpretation of it in terms of a local fibration.
Section 3 is devoted to defining the Poin\-ca\-r\'e--Ne\-kho\-ro\-shev map, which can now be
seen geometrically as a direct generalization of the standard Poincar\'e map;
the subsequent sections deal with applications of the Poin\-ca\-r\'e--Ne\-kho\-ro\-shev map.
In Section 4 we discuss the relation between fixed points of the Poin\-ca\-r\'e--Ne\-kho\-ro\-shev
map and invariant manifolds, in particular tori. Section 5 is devoted to
study persistence of invariant tori in non-hamiltonian systems from a
geometrical point of view; the same question is discussed in
Section 6 with the use of a coordinate system and thus providing
explicit formulas. The brief
Section 7 discusses invariant tori in hamiltonian systems
(i.e.\ the subject of \cite{Nek}) from the present standpoint. Finally, in
Section 8 we discuss how standard results for bifurcation of fixed
points of maps are to be interpreted in this frame as describing
bifurcations from an invariant torus.

\section{Notation and background}

We consider a smooth $n$-dimensional manifold $M$ (by smooth we will always
mean $C^r$  with some fixed $r$, $1 \le r \le \infty$, constant throughout the
paper), and in this $k$
independent smooth
 vector fields $X_i$, $i=1,\ldots,k$, spanning a
$k$-dimensional Lie algebra ${\mathcal G}$.
 We are specially interested in the case --
and thus we assume -- that ${\mathcal G}$ is
 abelian, i.e.\ $[ X_i , X_j ] = 0$.

We denote by $G$ the connected Lie group generated by ${\mathcal G}$, and by $G_0 = \{
\exp[ \varepsilon X] , \, X \in {\mathcal G} , \, - \varepsilon_0 < \varepsilon < \varepsilon_0 \} \subset G$ the {\it
local} Lie group generated by ${\mathcal G}$; local Lie groups are
 discussed e.g.\ in~\cite{DuK,Olv}.

We stress that in general $G$ is not compact, and we are {\it not} assuming
it acts regularly or with regularly embedded orbits in $M$.

Suppose now that there is a smooth
compact and connected
 submanifold $\Lambda \subset M$, which is ${\mathcal G}$-invariant (this
means $X_i : \Lambda \to {\mathrm T} \Lambda$ for all $i=1,\ldots,k$), and such that
 the $X_i$
are linearly independent at all points $m \in \Lambda$. As
 ${\mathcal G}$ is abelian and
$\Lambda$ is compact and connected, necessarily
 $\Lambda \simeq \toro^k$ (the
equivalence being a smooth isomorphism),
 see e.g.~\cite{Arn1}. Note also that
the linear independence of
 the $X_i$ at all points of $\Lambda$ implies that they
are linearly
 independent in a tubular neighbourhood $U \subset M$ of $\Lambda$.

As the $X_i$ are independent on $\Lambda$, we can choose coordinates
$(\varphi_1 ,\ldots, \varphi_k )$ on $\Lambda$ (with $\varphi_i \in S^1$) such
that $Y_i := (\partial / \partial \varphi_i)$, and the loops $\Gamma_i$
corresponding to the $\varphi_i$ coordinate running from $0$ to $2
\pi$ while the others remain constant can be chosen as basis cycles in $\Lambda$.

The homotopy class of a loop $\gamma$, which we will denote as
$h (\gamma) = \alpha = (\alpha_1,\ldots,\alpha_k) \in \Z^k$
(we also write $h_i (\gamma) = \alpha_i$), counts the winding of $\gamma$
around the basis cycles of $\Lambda$; with the choices mentioned
above and this notation,  $\alpha_i = h_i (\gamma)$ is just the increase of
$\varphi_i/(2 \pi)$ along the path $\gamma$.

The following lemma is well known, but we will however give a
proof of it, also in order to fix some notation.

\newpage

\begin{lemma}
Take any loop $\wt\gamma$ and any point $m
\in \Lambda$. Then there is a loop $\gamma$ with $h (\wt{\gamma}) = h (\gamma)$,
which is the orbit through $m$ of a vector field of the form
\begin{equation}
 X_\alpha = \sum_{i=1}^k c_i (\alpha) X_i
\end{equation}
 with suitable $c_i (\alpha) \in \R$.
\end{lemma}

\begin{proof}
The flow on $\Lambda$ under the vector field
\begin{equation}
 X_\alpha = 2 \pi \sum_{i=1}^k \alpha_i \, X_i
\end{equation}
 obeys the
equations $ (d \varphi_i / d s ) = 2 \pi \alpha_i$; these obviously have
the solution
\begin{equation}
 \varphi_i (s) = \varphi_i (0) + 2 \pi  \alpha_i s .
\end{equation}
For $s \in [0,1]$ this describes a loop $\gamma$ in
$\Lambda$ with homotopy class
$h (\gamma) = \alpha$. Thus for a~given path $\wt\gamma$ with
homotopy class $h (\wt\gamma ) = \alpha$, equation (2)
gives the required vector field, and equation (3) yields the homotopically
equivalent path $\gamma$, mentioned in the statement. Obviously there
is such a path through any point $m_0 = (\varphi_1 (0) , \ldots , \varphi_k
(0) ) \in \Lambda$. Note we have also determined the coefficients $c_i
(\alpha)$ appearing in (1): they are just $c_i (\alpha) = 2 \pi \alpha_i$.
\end{proof}

The notion of {\it monodromy operators} is
also well known, but we will quickly recall it, again in order to fix some notation.

Let $X$ be a vector field in $M$, and assume there is a nontrivial
closed orbit $\gamma$ for $X$ passing through $m \in M$ and having finite
period $\tau$; obviously we can always take $\tau = 1$ by rescaling $t$ or $X$, or both.

We consider then the {\it total monodromy map} $T_X = \exp [\tau X
]$, which maps the point $x = x (0) \in M$ to the point $x(\tau)$
on the flow $x(t)$ of $X$ with $x(0) = x$ at time $\tau$. With our
assumption, $T_X (m) = m$ for all points $m \in \gamma \subset M$.

Let us denote by $A_{(X,m)}$ the linearization
of $T_X$ at $m \in \gamma$, i.e.\ $ A_{(X,m)}  :=  [D T_X ]_m $. The $A_{(X,m)}$
is also called the {\it total monodromy operator}.

Note that, as the $X$ flow through $m$ is periodic,  $A_{(X,m)}$
will always have an eigenspace tangent to $X(m) \in {\mathrm T}_m M$ and
corresponding to an eigenvalue one. We can thus -- with no
loss of information -- consider the projection of $T$ and $A$
(denoted by $P$ and $L$ respectively) to a subspace transversal to
$X(m)$ in ${\mathrm T}_m M$; we will call them {\it transversal
monodromy map} and {\it transversal monodromy operator}. (Note this
terminology is not standard: in part of the literature these are defined to be
the monodromy map and
 monodromy operator; the reader will easily avoid
confusion by looking at the dimension of spaces involved.)

The eigenvalues $\lambda_i$ of $L_{(X,m)}$ are called
{\it characteristic (or Floquet) multipliers}; they carry most of the information needed
to study the dynamics defined by $X$ around the periodic orbit
$\gamma$.

It is also customary to write $L_{(X,m)} = \exp[\tau Q_{(X,m)}]$;
the eigenvalues $\chi_i$ of $Q_{(X,m)}$, having the
obvious relation $\lambda_i = \exp[\tau \chi_i]$ with those of $L_{(X,m)}$,
are called, {\it characteristic (or Floquet) exponents}.

\begin{remark} It is well known that monodromy operators at
different points in $\gamma$ are conjugated, so that the spectrum of
$A_{(X,m)}$ depends on $\gamma$ but not on the point $m \in \gamma$.
Similarly,  the monodromy operator based at $m \in \gamma$ around a
path $\gamma$ depends only on the homotopy class of $\gamma$, and not on
the actual path. See \cite{Arn2,Dem,Rue} for details.
\end{remark}

It will follow from this remark that in our subsequent discussion
-- where we will need only the spectrum of monodromy operators --
we can limit to consider vector fields of the form (1) and
correspondingly paths of the form (3), which also sets $\tau=1$;
moreover it is easy to see that different $X_\alpha$ orbits in $\Lambda$
are conjugated by a $G$-action. Note also that the monodromy map and
operator are invariant under a rescaling of $X$ and/or $t$
(recall $\tau$ is the period and hence changes accordingly). Thus
when dealing with $X_\alpha$ we will simply write $T_\alpha$, $A_\alpha$ and
$Q_\alpha$, and call $T_\alpha$ the ``time-one map'' under $X_\alpha$.

\section{The Poincar\'e map}

Let $M$ be a $n$-dimensional smooth manifold. As well known, the
Poincar\'e map is defined in the neighbourhood of (an arbitrary
point on) a periodic orbit of the vector field. Let $\gamma$ be a
nontrivial closed orbit through the point $m \in M$ for a smooth
vector field $X$. Consider a local manifold $\Sigma$ through $m$,
transversal to $\gamma$ in $m$ (it is well known that the Poincar\'e
map does not depend on our choice of $\Sigma$, see e.g.~\cite{Rue}).

For $\Sigma_0$ a suitably small neighbourhood of $m$
in $\Sigma$, orbits through points $x \in \Sigma_0 \subseteq \Sigma$, $x \not= m$, first
intersect $\Sigma$ at a point $x'$, in general with $x' \not= x$;
the Poincar\'e map $P$ is then defined as $P(x) = x'$
\cite{Arn2,Dem,Gle,GuH,Rue,Ver}. (We need the restriction to
$\Sigma_0 \subseteq \Sigma$ as the orbit through $x$ could fail to meet again $\Sigma$ if $x$ is too far from $m$.)

For the present discussion it will be convenient to define the
Poincar\'e map in a more geometric way (see also \cite{Dem}). We
put again the period of the periodic orbit $\gamma$ for the vector field
$X$ equal to one (just rescale $X$ or $t$ if needed).

Consider a ${\mathcal G}$-invariant neighbourhood $U$ of $m$ in $M$ (note
that a $G$-invariant neighbourhood could fail to exist). As $X(m)
\not= 0$, by the flow box theorem \cite{Arn2} we can choose in $U$
coordinates $(\xi^1, \ldots, \xi^n)$, say with $m=(0,\ldots,0)$, such
that $X = (\partial / \partial \xi^n)$.

We can, for the sake of simplicity, take $\Sigma$ to
be described by $\xi^n = 0$. Note that in this way --
or however identifying locally $\Sigma$ with its tangent space at $m$,
$S \subset {\mathrm T}_m M$ -- the Poincar\'e map can also
be thought as an application between (open sets in) linear spaces.

Consider then the time-one flow $x \mapsto T(x)$ of points in
$\Sigma_0$ under $X$. The point $m$ is obviously mapped again, by
construction, to itself: $T (m) = m$. Nearby points
$x=(\xi^1,\ldots,\xi^{n-1},0)$ are in general not mapped to
themselves, and not even mapped back to $\Sigma$. Let $\pi:U \to \Sigma$
be the projection operator to $\Sigma$, given in the $\xi$ coordinates by
$ \pi (\xi^1 , \xi^2 , \ldots, \xi^n) =
(\xi^1 ,\ldots, \xi^{n-1},0)$; it is clear that the Poincar\'e
map $P$ is described by
\[
P (x) = \pi [ T (x) ] .
\]

\begin{remark}
 In abstract terms, this description can be
reformulated as follows. The quotient by the $X$ action is well
defined in $U$; the Poincar\'e map is nothing else than the
time-one flow map under $X$, modulo this quotient.
 This can also be seen as introducing in $U$ the structure of a fiber
bundle $(\pi:U\to \Sigma)$ over $\Sigma$; the vector field $X$ is vertical, and
the Poincar\'e map is the projection to the base space of the time-one
flow under $X$.
\end{remark}

Let $L$ be the linearization of the Poincar\'e map at the fixed
point $m$, $L = (DP)(m)$. This is actually the transversal
monodromy operator and, as well known and mentioned above, the
spectrum of $L$ and that of the complete monodromy operator $A =
(DT)(m)$ are closely related. Indeed, let $\{ \lambda_1 , \ldots ,
\lambda_{n-1} \}$ be the eigenvalues of $L$; then the eigenvalues of
$A$ are $\{\lambda_1 , \ldots, \lambda_{n-1} ; \lambda_n = 1 \}$. The last
eigenvalue $\lambda_n = 1$ corresponds to the eigenspace spanned by $X
(m)$ in ${\mathrm T}_m M$ (i.e.~to the line $\xi^1 = \cdots = \xi^{n-1} = 0$)
and is always present, by construction, in the spectrum of~$A$.

\section{The Poincar\'e--Nekhoroshev map}

Let us now come back to consider the case of a ${\mathcal G}$-invariant
$\toro^k$ submanifold $\Lambda \subset M$ (as this is
closed and compact, ${\mathcal G}$-invariance implies $G$-invariance).
Choose a reference point $m \in \Lambda$, and a smooth local
submanifold $\Sigma \subset M$, transversal to $\Lambda$ in $m$.

Consider again a suitably small ${\mathcal G}$-invariant neighbourhood
$U \subseteq M$ of $m$ in $M$; as the commuting vector fields $X_i$
($i=1,\ldots,k$)
are nonzero and linearly independent in $m$, by the flow
box theorem we can choose local coordinates $(\xi^1 , \ldots , \xi^n)$
in $U$ such that the vector fields $X_i$ are written, in these coordinates, as $ X_i =
(\partial / \partial \xi^{r+i} )$, for $i=1,\ldots,k$; here and below, $r:=
n-k$. Again for ease of discussion, choose the $\Sigma$ to be
identified by $\xi^{r+1} = \cdots = \xi^n = 0$.

We denote now by $\pi: U \to \Sigma$
the operator of projection to $\Sigma$, given in the $(\xi^1,\ldots,\xi^r)$ coordinates by
\[
 \pi  (\xi^1 , \ldots, \xi^r, \xi^{r+1} , \ldots ,\xi^n ) = (\xi^{1} , \ldots ,\xi^r ; 0,\ldots,0 ) .
\]

The time-one flow under $X_\alpha$ will again define a local map $T_\alpha
: \Sigma_0 \to U$, where $\Sigma_0 \subseteq \Sigma$ is a suitable small
neighbourhood of $m$ in $\Sigma$.

\begin{definition} The {\it Poin\-ca\-r\'e--Ne\-kho\-ro\-shev} map ${\mathcal P}_{\alpha,m} :
\Sigma_0 \to \Sigma$ associated to the vector field $X_\alpha$ and based at $m
\in \Lambda$ is defined in this notation as $ {\mathcal P}_{\alpha,m} (x) = \pi [ T_\alpha (x)]$.
\end{definition}

Analogously to the standard Poincar\'e case, this description can
be reformulated in abstract terms.

The quotient by the ${\mathcal G}$ action is well defined in $U$;
the Poincar\'e map is nothing else than the time-one flow map under $X_\alpha$,
modulo this quotient.

This can also be seen again as introducing in $U$ the structure of
a fiber bundle $(\pi:U\to \Sigma)$ over $\Sigma$; the vector fields $X_i$
are vertical, and the Poincar\'e map ${\mathcal P}_{\alpha,m}$ is the
projection to the base space of the time-one flow under $X_\alpha$.

\begin{remark} The key point in this construction is that if
we consider a tubular neighbourhood ${\mathcal N}$ of $\Lambda$ in $M$ and
the Lie group $G$ generated by ${\mathcal G}$, the quotient ${\mathcal N} / G$
is ill-defined except in situations where the $G$ action
is known {\it apriori} to be pretty simple (regular orbits);
in this case a $G$-invariant neighbourhood is known to exist,
and moreover one can simply consider the quotient system and apply
on this the standard Poincar\'e theory. On the other hand, restricting to
a neighbourhood of the {\it local} smooth manifold $\Sigma_m$
we can always consider the quotient by the {\it local} Lie group $G_0$.
\end{remark}

\begin{remark} Let $Z$ be a contractible neighbourhood in
$\Lambda$, and define transversal manifolds~$\Sigma_m$ through any point
$m \in Z \subset \Lambda$; their union is a open set $W$. Inside this
there is a~${\mathcal G}$-invariant neighbourhood $W_0 \subseteq W$ of $Z$ which
can be seen as a trivial fiber bundle $(\mu:W_0 \to Z)$ over $Z$;
the vector fields $X_i$ are horizontal in $W_0$ and thus define a
field of horizontal $k$-planes, i.e.\ a connection in $W_0$. By
Frobenius' theorem, there are local smooth
$k$-manifolds in $W_0$ which are everywhere tangent to this field of $k$-planes and thus
${\mathcal G}$-invariant.
\end{remark}

\begin{remark} It should be stressed that the Poin\-ca\-r\'e--Ne\-kho\-ro\-shev map can
also be seen as the composition of two maps in a slightly
different way: time-one flow under $X_\alpha$ and the flow (for a time
$t_b (x)$ which we do not need to determine) under a vector
field $X_b = \sum_i b_i X_i$: indeed, any two points on the same
fiber $\pi^{-1} (x)$ can be joined in this way. It is immediate
from this that the Poin\-ca\-r\'e--Ne\-kho\-ro\-shev map is the composition of two smooth maps,
and is thus itself a smooth map.
\end{remark}

We will now consider the linearization $L_{\alpha,m}$ of the Poin\-ca\-r\'e--Ne\-kho\-ro\-shev map
${\mathcal P}_{\alpha;m}$ around the fixed point $m$; we are specially
interested in its spectrum.

It turns out that this spectrum is independent of the base point
$\varphi_0 \in \Lambda$, i.e.~depends only on the homotopy class $\alpha$;
moreover, it is simply related to the spectrum of the total
monodromy operator $A_{\alpha,m} := (DT_\alpha) (m)$ for $X_\alpha$.

It is clear that $A_{\alpha,m}$ always has $k$ eigenvalues equal to one;
these correspond to eigenspaces spanned by the $X_i$ (that is, tangent to $\Lambda$) at $m$.
In the $\xi$ coordinates, these span the subspace $\xi^{k+1} = \cdots = \xi^n = 0$.

This observation shows immediately the relation between the
spectra of $A_{\alpha,m}$ and of $L_{\alpha,m}$: if the spectrum of
$A_{\alpha,m}$ is given by $\{\lambda_1,\ldots,\lambda_r ; 1,\ldots,1 \}$
($r=n-k$), then the spectrum of $L_{\alpha,m}$ is given by $\{\lambda_1 ,
\ldots, \lambda_r \}$, and viceversa. (This of course also
establish a~relation between the spectra of $L_{\alpha,m}$ and that of
the transversal monodromy operator.)

\begin{lemma}
Given any two points $\varphi_0$ and $\varphi_1$
in $\Lambda$, and any homotopy class $\alpha \in \pi_1 (\Lambda)$, the
matrices $L_{\alpha;\varphi_0}$ and $L_{\alpha;\varphi_1}$ are conjugated; hence
their spectra coincide.
\end{lemma}

\begin{proof}
As recalled above, the spectra of monodromy
operators $A_{\alpha,m}$ only depend on $\alpha$, not on $m$; hence the
same holds for the spectra of the linearized Poin\-ca\-r\'e--Ne\-kho\-ro\-shev maps $L_{\alpha,m}$,
see above.
\end{proof}

\section{Invariant tori}

In this section we discuss -- to the extent needed for our goals
-- the relation between invariant tori and fixed points of the Poin\-ca\-r\'e--Ne\-kho\-ro\-shev
map.

We assume that there is a $G$-invariant submanifold ${\mathcal M}_0 \subseteq M$,
with $\Lambda \subseteq {\mathcal M}_0$; we denote by $L^{(0)}_{\alpha,m}$
and ${\mathcal P}^{(0)}_{\alpha,m}$ the restrictions of $L_{\alpha,m}$ and ${\mathcal P}_{\alpha,m}$ to ${\mathcal M}_0$.

\begin{lemma}
If $L_{\alpha,m}^{(0)}$ has no eigenvalue of
unit norm, then $\Lambda$ is an isolated $G$-invariant torus in
${\mathcal M}_0$.
\end{lemma}

\begin{proof}
Assume there is a $G$-invariant torus near $\Lambda$ in
${\mathcal M}_0$; it will intersect $\sigma_m$ in some point $x \not= m$
near $m$, and necessarily ${\mathcal P}_{\alpha,m}^{(0)} (x) = x$.
However, the condition on the spectrum of $L_{\alpha,m}^{(0)}$
implies it is an hyperbolic map, and thus the fixed point $m$
is isolated in $\sigma_m$. This in turn implies the lemma.
\end{proof}

Needless to say, if the spectrum of $L_{\alpha,m}$ (no restriction to
${\mathcal M}_0$) does not contain eigenvalues of unit norm, then $\Lambda$ is
isolated in $M$ and not just in ${\mathcal M}_0$.

The union of the stable and unstable manifolds for the torus $\Lambda$
is obviously ${\mathcal G}$-invariant. Then we have immediately from Lemma~3
the

\begin{corollary}
Let ${\mathcal H}$ be the union of the stable and
unstable manifolds for $\Lambda$; then there is no $G$-invariant torus
near $\Lambda$ in ${\mathcal H}$.
\end{corollary}

It is also obvious (and it has been used in the proof of Lemma~3
above) that a ${\mathcal G}$-in\-variant torus near $\Lambda$ corresponds to a
fixed point of the Poin\-ca\-r\'e--Ne\-kho\-ro\-shev map; let us discuss if fixed point of the
map correspond to invariant manifolds, and sufficient conditions
for these to be tori.

\begin{lemma}
Let $x \in \Sigma_m$ be a fixed point for the
Poin\-ca\-r\'e--Ne\-kho\-ro\-shev map ${\mathcal P}_{\alpha,m}$; then there is a ${\mathcal G}$-invariant smooth
manifold through $x$.
\end{lemma}

\begin{proof}
If $x=m$ the assertion is trivial, so assume $x
\not= m$. By Remark~4, for $Z$ a~neighbourhood of $m$ in $\Lambda$
there is a well defined local ${\mathcal G}$-invariant manifold $Y_0$ through
$x$; with the construction introduced there, call $y(p)$ the point
$Y_0 \cap \Sigma_p$, $p \in Z$ (so $y(m) = x$). Note that ${\mathcal
P}_{\alpha,m} [y(m)] = y(m)$ implies ${\mathcal P}_{\alpha,m} [y(p)] = y(p)$
for all $p \in Z$. Consider now an atlas~$\{ Z_i \}$ of $\Lambda$:
there is a local ${\mathcal G}$-invariant manifold $Y_i$ over each chart
$Z_i$, and by considering $Z_i \cap Z_j$ it is immediate to check
that the transition functions are also smooth.
Hence the $Y_i$ blend together to give a smooth manifold $Y$, ${\mathcal G}$-invariant by construction.
\end{proof}

Let us now consider the case where there is a $G$-invariant
submanifold ${\mathcal M}_\beta \subset M$; define $\sigma^{(\beta)}_m := \Sigma_m \cap
{\mathcal M}_\beta$, and note that ${\mathcal P} [ \sigma^{(\beta)}_m ] \cap \Sigma_m \subseteq
\sigma^{(\beta)}_m$. We can thus define the restriction of the Poin\-ca\-r\'e--Ne\-kho\-ro\-shev map to
$\sigma^{(\beta)}_m$, denoted as ${\mathcal P}^{(\beta)}$.

\begin{lemma}
Let ${\mathcal M}_\beta$, $\sigma^{(\beta)}_m$ and ${\mathcal
P}^{(\beta)}_m$ be as above, and let $x \in \sigma_m$ be the unique fixed
point for ${\mathcal P}^{(\beta)}_m$. Then there is a $G$-invariant
$k$-torus through $x$, smoothly equivalent to $\Lambda$.
\end{lemma}

\begin{proof}
This follows immediately from the construction
used in previous lemma and the unicity of $x$: in this case there
is a smooth one-to-one correspondence between points of~$\Lambda$ and
points on the ${\mathcal G}$-invariant manifold $Y(x)$.
As this is closed and compact, it is also $G$-invariant.
\end{proof}

\section{Persistence of invariant tori}

We want to consider the case where the vector fields $X_i$ depend
smoothly on parameters; we aim at local results in the parameter
space, so we will denote these parameters as $\varepsilon \in {\mathcal E} \subseteq E =
\R^p$, and write $X_i^{(\varepsilon)}$. In this case we deal with a
smooth manifold ${\mathcal M} = {\mathcal E} \times M$, which is foliated into
$G$-invariant smooth submanifolds ${\mathcal M}_\varepsilon = \{ \varepsilon \} \times M
\simeq M$.

We assume $\Lambda \equiv \Lambda_0$ is an invariant
torus for all the $X_i^{(0)}$ and wonder if -- and under which conditions
-- this persists under perturbation, i.e.~if there is some
torus $\Lambda_\varepsilon$, near to $\Lambda_0$ and
invariant under all the $X_i^{(\varepsilon)}$, for $\varepsilon \not= 0$ small enough.

Let us recall what is the situation for $k=1$, i.e.~for a single vector field $X^{(\varepsilon)}$ and
a~periodic orbit $\gamma_0$ of the vector field $X^{(0)}$. It is well known that,
with an obvious extension of the notation considered in Section~2,
the {\it Poincar\'e--Lyapounov theorem}
states that if the eigenvalues $\lambda_i$ of the transverse monodromy
operator $L^{(0)}$ associated to the path $\gamma$ satisfy $|\lambda_i | \not= 1$,
then the periodic orbit $\gamma$ is
actually part of a continuous $p$-parameters family of periodic
orbits for $X$. This amounts essentially to using the implicit
function theorem (see \cite{Arn2,Rue} or e.g.~\cite{AmP}, or any
text in nonlinear analysis) for the Poincar\'e map, in order to
ensure there is a $p$-parameters branch of fixed points for it,
and recognizing that fixed points of the Poincar\'e map
corresponds to periodic orbits.

A similar result, the {\bf Poincar\'e--Lyapounov--Nekhoroshev
theorem}, was obtained by Ne\-kho\-ro\-shev \cite{Nek} in the case of
invariant tori, in terms of the spectra of the Poin\-ca\-r\'e--Ne\-kho\-ro\-shev maps associated
to a generating set of homology cycles for the torus $\Lambda$.
Although his formulation was for hamiltonian dynamical
systems, the theorem holds --
with simple modifications -- for general  ones, and we will discuss
it in this general setting (see Section~7 below for the hamiltonian case).

\begin{theorem}[Nekhoroshev]
Let $M$ be a $n$-dimensional smooth manifold,
and ${\mathcal E} = E_0$ a neighbourhood of the origin in
 $E = \R^p$. Let $X_1^{(\varepsilon)},\ldots,X_k^{(\varepsilon)}$
be $k$ smooth vector fields on $M$ ($1 \le k \le n$),
smoothly dependent on the $p$-dimensional parameter $\varepsilon \in {\mathcal E}$,
independent for all $\varepsilon \in {\mathcal E}$, and such that $[X_i^{(\varepsilon)} , X_j^{(\varepsilon)} ] = 0 $
for all $\varepsilon \in {\mathcal E}$. We write ${\mathcal M} = {\mathcal E} \times M$, ${\mathcal M}_\varepsilon = \{ \varepsilon \} \times M$,
and denote by ${\mathcal G}^{(\varepsilon)}$ the Lie algebra spanned by the $X_i^{(\varepsilon)}$.

Assume that:

{\rm (i)} there exists a smooth $k$-dimensional torus $\Lambda_0 \subset
{\mathcal M}_0$ invariant under all the $X_i^{(0)}$, and that these are
linearly independent at all points of $\Lambda_0$;

{\rm (ii)} there is a $c \in R^k$ such that the vector field
$X_c^{(0)} = \sum_i c_i X_i^{(0)}$ has nontrivial closed
trajectories with finite period $\tau$ in $\Lambda_0$;

{\rm (iii)} the spectrum of the linear part $L_c^{(0)}$ of the
Poincar\'e-Nekhoroshev map associated to $X_c^{(0)}$ lies at a
distance $\delta > 0$ from the unity.

Then, in a neighbourhood ${\mathcal V}$ of $\Lambda_0$ in ${\mathcal M}$,
there is a smooth submanifold ${\mathcal N} \subset {\mathcal V} \subset {\mathcal M}$
which is fibered over the domain ${\mathcal E}$ with as fibers smooth tori $\Lambda_\varepsilon \simeq \toro^k$,
smoothly equivalent to $\Lambda_0$ and ${\mathcal G}^{(\varepsilon)}$-invariant.
\end{theorem}

\begin{proof}
We will focus on a point $m \in \Lambda_0$;
choose a smooth submanifold $\Sigma \subset {\mathcal M}$
transversal to $\Lambda_0$ in $m$. By choosing suitable
coordinates -- basically, those of the tangent space ${\mathrm T}_m \Sigma \subset {\mathrm T}_m {\mathcal M}$ --
we can identify a neighbourhood $\Sigma_0$ of $m$ in $\Sigma$ to a neighbourhood
$S_0$ of the origin in a linear space $S$.

We define the submanifolds $\sigma^{(\varepsilon)} := \Sigma \cap {\mathcal M}_\varepsilon$,
and let $\sigma^{(\varepsilon)}_0 := \Sigma_0 \cap {\mathcal M}_\varepsilon$. In the same way
as $\Sigma_0$ can be identified with a neighbourhood $S_0$ of the
origin in the linear space $S$, the manifold $\sigma^{(\varepsilon)}_0$ can
 be identified with a neighbourhood $U^{(\varepsilon)}_0$ of the origin in a linear space
$U_\varepsilon = \{ \varepsilon \} \times U \simeq U \subset S$.

As the $X_i^{(\varepsilon)}$ do not act on the value of $\varepsilon$, the
submanifolds ${\mathcal M}_\varepsilon$ are trivially $G$-invariant, and by
construction ${\mathcal P}_{\alpha,m} : \sigma^{(\varepsilon)}_0 \to \sigma^{(\varepsilon)}$.
 We denote by ${\mathcal P}^{(\varepsilon)}_{\alpha,m}$ the restriction of ${\mathcal P}_{\alpha,m}$ to $\sigma^{(\varepsilon)}_0$.

It will also be convenient to separate the coordinates in ${\mathcal E}$ and
those in $U \simeq U_\varepsilon$: a~point $x \in S_0$ will be denoted by
coordinates $(\varepsilon , u) \in {\mathcal E}_0 \times U_0 \subset E \times U = S$.
Thus we have coordinates $(\varphi,u;\varepsilon)$ with $\varphi \in \toro^k$,
$u \in U_0 \subset \R^{(n-k)}$, and $\varepsilon \in E_0 \subset \R^p$. We will write $N = k + r +p = n+p$.

By Lemma 1, we can consider $X_\alpha$ rather than
$X_c$, where $\pi_1 (c) = \alpha$. The discussion of Section 4 shows
that the theorem can be restated in terms of -- and proved by
studying~-- fixed points of the Poin\-ca\-r\'e--Ne\-kho\-ro\-shev map ${\mathcal P}_{\alpha,m}$
associated to $\alpha$ and based at an arbitrary point $m$.
We will think of $\alpha$ and $m$ as fixed and omit indices
 referring to these, for ease of notation.

As remarked above, we can actually consider the restrictions of
the Poin\-ca\-r\'e--Ne\-kho\-ro\-shev map to the submanifolds $\sigma^{(\varepsilon)}$; we will thus look
for fixed points of ${\mathcal P}^{(\varepsilon)}$.

Actually, it is convenient to slightly modify this formulation:
considering the map $\Phi : S_0 \to S$ defined by
\[
 \Phi (x) := x - {\mathcal P} (x) ,
\]
fixed points of the Poin\-ca\-r\'e--Ne\-kho\-ro\-shev map correspond to zeroes of $\Phi$, and we know that $\Phi (m) = 0$.
Passing to the $(\varepsilon,u)$ coordinates, we deal with a smooth
map $\Psi : E_0 \times U_0 \to E \times U$, defined by
$\Psi (\varepsilon,u) := ( \varepsilon , u - {\mathcal P}^{(\varepsilon)} (u) )$, and its restriction to $U_\varepsilon$ is therefore
\[
 \Psi^{(\varepsilon)} (u)  :=  \big( u - {\mathcal P}^{(\varepsilon)} (u) \big) .
\]

Consider the $r$-dimensional linear operator ${\mathcal B} : U_0 \to U$ defined as
\[
 {\mathcal B}  := \big(D_u \Psi^{(0)}\big)_m  = I - L^{(0)} .
\]
By the implicit function theorem (see e.g.~\cite{AmP}),
 if $\Psi^{(0)} (u_0) = 0$ and ${\mathcal B}$ is invertible,
then there are neighbourhoods $\wt{E} \subset E_0$ of $\varepsilon_0=0$
and $\wt{U} \subset U_0$ of $u_0=0$, and a smooth map $g : \wt{E} \to \wt{U}$,
such that $\Psi(\varepsilon,g(\varepsilon))  =  0 $ for all $\varepsilon \in \wt{E}$;
and moreover $\Psi(\varepsilon,u) =  0$ with $(\varepsilon,u) \in \wt{E} \times \wt{U}$ implies $u = g(\varepsilon)$.

In other words, if ${\mathcal B} : U_0 \to U$ is invertible, then there is a
unique fixed point of ${\mathcal P}$, i.e.~$(\varepsilon , g (\varepsilon ))$,
on each $\sigma^{(\varepsilon)}$, for $\varepsilon \in \wt{E} = E_0$.

Due to the results of Section 4, this means that there is an
invariant torus $\Lambda_\varepsilon \simeq \toro^k$, $\Lambda_\varepsilon \subset {\mathcal M}_\varepsilon$,
for each $\varepsilon \in \wt{E}$, provided ${\mathcal B}$ is invertible. However, the
eigenvalues $\beta_i$ of ${\mathcal B}$ and $\lambda_i$ of $L^{(0)}$
are simply related by $ \beta_i = 1 - \lambda_i $, so ${\mathcal B}$ is
invertible provided $\lambda_i \not= 1$ for all $i=1,\ldots,n$. This concludes the proof.
\end{proof}

\begin{remark} Note that it is sufficient that there is one
closed path with associated monodromy operator satisfying the
condition (iii) of the theorem to ensure persistence of invariant
tori.
\end{remark}

\begin{remark} The theorem and its proof are immediately
generalized to the case of infinite dimensions; see~\cite{BV}
for the hamiltonian case and an application to breathers.
\end{remark}

\begin{remark} If we consider the general case of ${\mathcal M}$ a $N$-dimensional
smooth manifold, ${\mathcal E} \subset {\mathcal M}$ a $p$-dimensional manifold, and
 assume ${\mathcal M}$ is foliated by regular $G$-invariant submanifolds
 ${\mathcal M}_\varepsilon$ ($\varepsilon \in {\mathcal E}$), with $\Lambda_0 \in {\mathcal M}_0$, the theorem
 remains true. Indeed, our construction is purely local and is still
valid (with $E$ the tangent space to ${\mathcal E}$ in $\varepsilon_0 = 0$, $M \simeq {\mathcal M}_0$).
\end{remark}

\section{The coordinate approach}

So far our discussion has been mainly geometrical; in this
section we will translate it into explicit formulas, making
use of the $(\varphi,u;\varepsilon)$ coordinates defined above
 (recall that $\varphi \in \toro^k$, $u \in U_0 \subset U = \R^r$,
$\varepsilon \in E_0 \subset E = \R^p$) in a neighbourhood
${\mathcal V} \subset {\mathcal M}$ of $\Lambda$. We stress that we do {\it not}
 assume ${\mathcal V}$ is $G$-invariant, nor we use ${\mathcal G}$-adapted coordinates.

As the vector fields $X_i \equiv X_i^{(\varepsilon)}$ do not act on the
parameters $\varepsilon$, their expression in these coordinates will be
\[ X_i =
\sum_{j=1}^k \, f_i^{~j} (\varphi,u;\varepsilon) \, \frac{\partial}{\partial \varphi_j}
+  \sum_{\mu=1}^r \, F_i^{~\mu} (\varphi,u;\varepsilon) \, \frac{\partial}{\partial
u_\mu} \, .
\]

For ease of notation, from now on summation
over repeated indices will be tacitly understood;
latin indices other than $\ell$ will run from 1 to $k$, while
$\ell = 1,\ldots,p$, and greek indices will run from 1 to $r$.
We also write $ \partial_i := (\partial / \partial \varphi^i )$ and $\partial_\mu  :=  (\partial / \partial u^\mu )$.

The torus $\Lambda \equiv \Lambda_0$ corresponds to $u=0,\varepsilon=0$, and its
invariance guarantees the vanishing of $ F_i^{~m} (\varphi,0;0)$.
Similarly, in Section 1 the coordinates $\varphi$ were chosen so
that on $\Lambda_0$ we had $X_i = (\partial / \partial \varphi^i )$; hence
$ f_i^{~j} (\varphi,0;0) = \delta_i^{~j}$.

Expanding $X_i$ at first order in $\varepsilon$ and $u$
around $\Lambda$, and considering then $X_\alpha = c^i (\alpha) X_i$,  we get
\begin{equation}
 X_\alpha  =  c^i \partial_i  +  \big[ \wt{P}^j_\nu u^\nu + \wt{Q}^j_\ell \varepsilon^\ell \, \big] \partial_j
 +  \big[ \wt{A}^\mu_\nu u^\nu + \wt{B}^\mu_\ell \varepsilon^\ell \,\big] \partial_\mu + {\rm h.o.t.} ,
\end{equation}
where $h.o.t.$ denotes higher order terms in $(u,\varepsilon)$, $c^i \equiv c^i (\alpha)$, and
 \begin{gather}
\wt{P}^j_\mu  :=  c^i \, (\partial f_i^j \, / \, \partial u^\mu) ; \qquad
\wt{Q}^j_\ell  :=  c^i \, (\partial f_i^j \, / \, \partial \varepsilon^\ell) ; \nonumber\\
\wt{A}^\mu_\nu := c^i \, (\partial \psi_i^\mu \, / \, \partial u^\nu) ; \qquad
\wt{B}^\mu_\ell  :=  c^i \, (\partial \psi_i^\mu \, / \, \partial \varepsilon^\ell)  .
 \end{gather}
All partial derivatives are computed on $\Lambda$, so that the matrices
 $\wt{A}$, $\wt{B}$, $\wt{P}$, $\wt{Q}$ are function of $\varphi \in \toro^k$.

We write $\varphi^i (t) = \varphi^i_0 (t) + \vartheta (t)$, where $\varphi^i_0 (t) = \varphi^i (0) + c^i t $
and $\vartheta \simeq O(\varepsilon,u)$. Therefore, as we keep only first order terms
in $(\varepsilon,u)$ in the expression for $X_\alpha$, see (4),
we can consider $\wt{A} (\varphi ) \simeq \wt{A} (\varphi^i_0 (t)) := \^A (t)$,
and similarly for the other matrices. Note that $\^A$ and
the like are explicit periodic functions of time.

The linearized flow around $\Lambda$ under $X_\alpha$ is hence described by
\begin{equation}
 \dot{u}   =  \^A u  +  \^B \varepsilon  ; \qquad
\dot{\vartheta}  =  \^P u  +  \^Q \varepsilon  ; \qquad
\dot{\varepsilon}  = 0 .
\end{equation}
However, for the sake of discussing the Poin\-ca\-r\'e--Ne\-kho\-ro\-shev map only the first equation is relevant.

Our discussion in the previous section shows that
we can actually consider just the restriction of this dynamics
to the space $\varepsilon=0$, in which case we just deal with
\begin{equation}
 \dot{u}  =  \^A (t) u  ,
\end{equation}
i.e.~a linear ODE in $\R^r$ with periodic coefficients.
The method of analysis of such equations is well known (see e.g.~\cite{Gle,Ver}),
and we briefly recall it.

One considers a fundamental matrix for $\^P(t)$ (this is
 built with a set of $r$ independent solutions); this matrix
$\Theta (t)$ satisfies $\dot{\Theta} = \^A \Theta$. By
Floquet's theorem~\cite{Ver}, it is always possible to write
$ \Theta (t) = M (t) \, \exp [B t] $, with $M$ a periodic and $B$ a constant matrix.
Then one performs the change of variables $ u = M (t)\, v $;
using $\dot{\Theta} = \^A \Theta$, and thus
$\dot{M} = (\^A M - M B)$, and the existence of $M^{-1}$, one gets $\dot{v} = B v$.

With a matrix $R$ of eigenvectors for $B$,
we can further write $v = R w$, and get $ \dot{w}  =  D w$,
where $D = R^{-1} B R = {\rm diag} (\lambda_1 ,\ldots , \lambda_r )$.
The solution of this is obviously $w^i(t) = \exp[ \lambda_i t] w^i (0)$
(no sum on $i$), which yields $ u (t) = M (t) R (\exp[ D t]) (R^{-1})  (M^{-1}) (0) \, u (0) $.

At time $t=T$ we get $ u(T) = Q u(0)$, where
$Q = M_0 R (\exp [DT]) R^{-1} M_0^{-1}$, and $M_0 = M(0)=M(T)$.
By definition, $Q$ is the monodromy matrix for (7),
and obviously the spectrum of $Q$ -- which is
the same as that of $B$ -- is just given by $\mu_i := \exp [\lambda_i T]$.
The $\mu_i$ are the characteristic multipliers, and the $\lambda_i$ are
the characteristic exponents, for (7).

\begin{remark}
Note that the situation is rather different if we want to compute
the Floquet exponent for $\varepsilon \not= 0$: indeed in this
case we deal with an equation of the form
$ \dot{u} = A u + b $,
with $A = A(t)$ a periodic matrix and $b=b(t)$
a periodic vector function, $b^\mu = \^B^\mu_\ell \varepsilon^\ell$,
see~(6). Proceeding as above we arrive at
\[
 \dot{w}  =  D \, w  +  f(t) , \qquad D = {\rm diag} (\lambda_1 ,\ldots , \lambda_r ) ;
\]
here $f(t)$, obtained by the action of periodic matrices on periodic vectors,
is still periodic with the same period $T$.

If the periods of small $u$ oscillations for $\varepsilon=0$
are different from $T$, i.e.~if the cha\-rac\-teristic multipliers $\mu_i$
computed above satisfy $\mu_i \not= 1$, the
solution will be of the form $w^i(t) = \exp[ \lambda_i t] w^i (0) + F^i (t)$
 (no sum on $i$) with $F^i$ a periodic function; this does
not affect the period maps and the discussion remain valid with the
same monodromy matrix~$Q$.

On the other hand, if there is some characteristic
multiplier $\mu_i = 1$, solutions will not be of the same form,
and terms proportional e.g.\ to $t \exp[ \lambda_i t]$ will appear.
\end{remark}

\begin{remark}
It should be stressed that, as clear from the discussion in this section,
all we need to know in order to ensure the conditions of the theorem
are satisfied are the matrices of partial derivatives
$ (\partial \psi_i^\mu /  \partial u^\nu)$ computed at $u=0$,
$\varepsilon=0$; they concurr to form~$\^A$, see~(5).

This could be understood in a slightly different way: write
 $z = (u,\varepsilon) \in \R^{r+p}$; then the linearized evolution
equations for $z$ read $\dot{z} = W z$, with
\[
 W  =  \begin{pmatrix} \^A & \^B \\ 0 & 0 \end{pmatrix} ;
\]
the spectrum of $W$ is given by $\lambda=0$ (with multiplicity $p$)
and by the eigenvalues $\lambda_1 , \ldots , \lambda_r$ of $\^A$.
Thus we just have to check these satisfy $\lambda_i \not= 1$.
\end{remark}

\section{The hamiltonian case}

In the hamiltonian case, we consider a symplectic manifold $({\mathcal M}^{2n},\omega)$,
and $k$ independent and mutually commuting hamiltonians $H_1,\ldots,H_k$
(commutation is meant, of course, with respect to the Poisson bracket $\{\cdot,
\cdot\}$ defined by the symplectic form $\omega$). Each of these defines
a (hamiltonian) vector field $X_i$ by $i_X (\omega) = {\rm d} H_i$,
and $\{ H_i , H_j \} = 0$ implies $[X_i , X_j ]=0$.
We denote by ${\mathcal G}$ the abelian Lie algebra spanned by the $X_i$, and by $G$ its Lie group.

Note that the $(H_1,\ldots,H_k)$ are common constants of
motion for any dynamics defined by a linear combination of the
$H_i$ (equivalently, of the vector fields $X_i$), so that their values $(h^1,\ldots,h^k)$
can be seen as parameters. We denote the common level manifold
$H_i = h^i$ by ${\bf H}^{-1} (h)$.

If there exists a ${\mathcal G}$-invariant torus $\Lambda_0 \subset {\mathcal M}$,
 his is necessarily contained in ${\bf H}^{-1} (h_0) $
for some $h_0 \in \R^k$; we write then $h^i = h^i_0 + \varepsilon^i$.
We also assume the $X_i$ are independent on $\Lambda_0$.

We are thus exactly in the scheme discussed in previous sections, with $p=k$.

\begin{remark}
Note that, using freely the notation introduced above,
the variables canonically conjugated to the $(\varphi^1,\ldots,\varphi^k)$
via the symplectic structure are proportional to the $(\varepsilon^1,\ldots,\varepsilon^k)$;
this also implies that the characeristic multipliers relative to eigenvectors
in the space ${\mathrm T}_\Lambda E \subset {\mathrm T}_\Lambda {\mathcal M}$ are the same as those relative
to eigenvectors in ${\mathrm T}_\Lambda \Lambda \subset {\mathrm T}_\Lambda {\mathcal M}$, i.e.~are all equal to one.

However, this is no problem as far as Nekhoroshev
theorem is concerned: the eigenvalues relative to the parameter space
do not affect the spectrum of the operator ${\mathcal B}$, see Section~5.
\end{remark}

Actually Nekhoroshev's result \cite{Nek} also include
a second part, also referred to as the {\it Liouville--Arnold--Nekhoroshev theorem},
concerning the possibility of defining action-angle coordinates
in the symplectic submanifold $N \subset M$ fibered by invariant
isotropic tori; needless to say this second part is purely hamiltonian.
Note that here we need that all monodromy operators associated to basis
 cycles are to be nondegenerate in the sense of~$(iii)$ in order to be able
to extend action-angle coordinates (compare with Remark~6).

For a detailed discussion -- and proof -- of the Poincar\'e--Lyapounov--Nekho\-ro\-shev
theo\-rem in the hamiltonian case the reader is referred to~\cite{BG,Gae}.

\section{Bifurcation from an invariant torus}

The Poin\-ca\-r\'e--Ne\-kho\-ro\-shev map can be discussed in the same way as the standard Poincar\'e map
(or any map between open sets in real spaces); this includes in particular its
bifurcations when external parameters are varied. In this section we
illustrate the picture emerging from such a discussion when we deal with a
single parameter $\varepsilon \in {\mathcal E} \subseteq \R$ (thus $p =1$). Essentially we are just
interpreting the discussion of~\cite{Arn2} (section 34) on bifurcation of
fixed points of the Poincar\'e map in the present frame, so we will be rather
sketchy.

We assume that there is a fixed point $u_0 (\varepsilon) = 0$
for all values of $\varepsilon \in E_0 \subseteq {\mathcal E}$, stable for $\varepsilon < 0$
and loosing stability at $\varepsilon > 0$. This corresponds, for the
full dynamics, to a~(parameter-dependent) invariant torus
$\Lambda_0 (\varepsilon) = \toro^k$, which is transversally hyperbolically
stable for $\varepsilon < 0$ and looses stability for $\varepsilon > 0$. This implies
that some eigenvalues $\mu_i (\varepsilon)$ of the map ${\mathcal P}^{(\varepsilon)}$ satisfy $|\mu_i (0)| = 1$ .

Let us make standard bifurcation hypotheses, i.e.: $(i)$ the
existence of a dynamically invariant neighbourhood of $u_0$ for
all values of $\varepsilon \in E_0$; $(ii)$ trasversality for the
critical eigenvalues $\mu_i (\varepsilon)$, i.e.\ $d |\mu_i|/d
\varepsilon \not= 0$ at $\varepsilon = 0$; $(iii)$ non-degeneracy
of the spectrum of the map at the critical point (for generic
dynamics, this means that there is only a pair of complex
conjugate complex critical eigenvalues, or a single real one);
$(iv)$ split property of the spectrum: non-critical eigenvalues
lie at a finite distance $\delta > 0$ from the unit circle at
$\varepsilon = 0$.

With these, it is known that there are three elementary types
of bifurcation, characterized by the value of the critical eigenvalues
 $\mu_i (\varepsilon)$, i.e.\ of the eigenvalues $\mu_i$:
\begin{gather*}
(a) \ \ \mu (0) = -1 ;
\end{gather*}
\begin{gather*}
(b) \ \ \mu (0) = 1 ;
\\
(c) \ \ \mu (0) = \cos (\alpha) \pm i \sin (\alpha) \quad (\alpha \not= k \pi)  .
\end{gather*}

Case $(a)$ corresponds to the appearance of two new (branches of)
stable fixed points $u_\pm (\varepsilon)$ for the map; these corresponds
to a bifurcation of the invariant torus $\Lambda_0 (\varepsilon) \simeq \toro^k$
into two new (branches of) stable tori $\Lambda_\pm (\varepsilon) \simeq \toro^k$.
For $\varepsilon \to 0^+$, $u_\pm (\varepsilon) \to u_0 (\varepsilon)$, and similarly $\Lambda_\pm (\varepsilon) \to \Lambda_0 (\varepsilon)$.

Case $(b)$ corresponds to the appearance of two period-two points $u_\pm (\varepsilon)$,
such that ${\mathcal P}^{(\varepsilon)} : u_\pm (\varepsilon) \to u_\mp (\varepsilon)$. This is a period-doubling
bifurcation, and corresponds to the appearance of a single invariant torus $\Lambda_d (\varepsilon)$.
For $\varepsilon \to 0^+$, $u_\pm (\varepsilon) \to u_0 (\varepsilon)$, and $\Lambda_d (\varepsilon ) \to \Lambda_0 (\varepsilon)$.
For $\varepsilon > 0$ sufficiently small, $\Lambda_d (\varepsilon)$ lies near enough to $\Lambda_0$ to
make sense to consider its intersection with the transversal local manifolds to $\Lambda_0$,
and it has two such intersections on each $\sigma^{(\varepsilon)}$, given indeed by $u_\pm (\varepsilon )$.

Case $(c)$ is the most interesting; we can consider $2 \pi / \alpha$ irrational,
as the rational case is structurally unstable (see e.g.\ the discussion
in~\cite{Arn2}). In this case we get a full circle of fixed points $u_\vartheta (\varepsilon)$ ($\vartheta \in S^1$)
for the Poin\-ca\-r\'e--Ne\-kho\-ro\-shev map. This corresponds to the appearance of a new
stable torus $\Lambda_1 (\varepsilon) \simeq \toro^{k+1}$, of dimension
greater than that of the original invariant torus. This is the analogue
of the bifurcation of an invariant torus off a~periodic solution.

\begin{remark}
If we discuss the (nongeneric) symmetric case, i.e.\
 if we assume that there is an algebra -- no matter if abelian or otherwise --
of vector fields ${\mathcal H}$ commuting with~${\mathcal G}$,
then the nondegeneracy assumption
$(iii)$ should be meant in the sense that only the degeneracy imposed by
the symmetry constraint is present in the critical spectrum, see~\cite{Ru0}.
In this case there will be a multiplicity if critical eigenvalues which in case $(c)$
can lead to a~bifurcation in which the new stable torus is $\Lambda_1 (\varepsilon) \simeq \toro^{k+s}$
with $s > 1$.
\end{remark}

\subsection*{Acknowledgements}
Common work with Dario Bambusi on the Poincar\'e--Lyapounov--Nekhoroshev
theorem in the  hamiltonian case is at the basis of this work; my interest
in the general case and the geometrical aspects discussed here was
also triggered by questions raised by Franco Cardin; I would like to warmly
thank both of them.

 This work was supported by ``{\it Fondazione CARIPLO per la ricerca
scientifica}'' under the project ``{\it Teoria delle perturbazioni  per sistemi con simmetria}''.

\label{gaeta-lastpage}
\end{document}